\documentclass[runningheads]{llncs}
\usepackage{graphicx}
\usepackage[english]{babel}
\usepackage[utf8]{inputenc}
\usepackage[caption=false,font=footnotesize]{subfig}
\usepackage{caption}
\usepackage{booktabs}
\usepackage{amsmath,amssymb}
\usepackage{hyperref}
\usepackage[table]{xcolor}
\usepackage{collcell}
\usepackage{hhline}
\usepackage{pgf}
\usepackage{hyperref}
\usepackage{blindtext}
\usepackage{enumitem}
\usepackage{wrapfig}
\usepackage{soul}
\usepackage[normalem]{ulem}

\usepackage{booktabs}
\usepackage{makecell} 
\usepackage{arydshln}
\usepackage{amsmath}
\usepackage[caption=false,font=footnotesize]{subfig}
\usepackage[export]{adjustbox}
\usepackage{xcolor}

\def\colorModel{hsb} %
\newcommand\ColCell[1]{
  \pgfmathparse{#1<50?1:0}  %
    \ifnum\pgfmathresult=0\relax\color{white}\fi
  \pgfmathsetmacro\compA{0}      %
  \pgfmathsetmacro\compB{#1/100} %
  \pgfmathsetmacro\compC{1}      %
  \edef\x{\noexpand\centering\noexpand\cellcolor[\colorModel]{\compA,\compB,\compC}}\x #1
  } 
\newcolumntype{E}{>{\collectcell\ColCell}m{0.45cm}<{\endcollectcell}}  %

\usepackage[misc]{ifsym}
\def\Plus{\texttt{+}}

\definecolor{darkgreen}{rgb}{0.0,0.40,0.0}
\newcommand{\g}[1]{\textcolor{black}{#1}}

\begin{document}
\title{Convolutional Nets Versus Vision Transformers for Diabetic Foot Ulcer Classification}
\author{Adrian Galdran\inst{1,2} \and
Gustavo Carneiro\inst{2} \and
Miguel A. Gonz\'alez Ballester\inst{1,3}}
\authorrunning{A. Galdran et al.}
\institute{BCN Medtech, Dept. of Information and Communication Technologies, \\Universitat Pompeu Fabra, Barcelona, Spain, \email{adrian.galdran@upf.edu} \and
Australian Institute for Machine Learning, University of Adelaide, Australia \and ICREA, Barcelona, Spain}
\maketitle              %
\begin{abstract}
This paper compares well-established Convolutional Neural Networks (CNNs) to recently introduced Vision Transformers for the task of Diabetic Foot Ulcer Classification, in the context of the DFUC 2021 Grand-Challenge, in which this work attained the first position.
Comprehensive experiments demonstrate that modern CNNs are still capable of outperforming Transformers in a low-data regime, likely owing to their ability for better exploiting spatial correlations. 
In addition, we empirically demonstrate that the recent Sharpness-Aware Minimization (SAM) optimization algorithm improves considerably the generalization capability of both kinds of models. 
Our results demonstrate that for this task, the combination of CNNs and the SAM optimization process results in superior performance than any other of the considered approaches.
\end{abstract}

\section{Introduction and Problem Statement}
Diabetes affects 425 million people with a projection to reach 629 million affected people \g{by 2045 \cite{cho_idf_2018}}. 
A particularly concerning development of diabetes are foot ulcers: \g{the risk of a diabetic patient developing a foot ulcer along their lifetime is estimated to be 19-34\% \cite{armstrong_diabetic_2017}}.

Diabetic Foot Ulcers (DFU) presented with ischaemia and/or infections can lead to limb gangrene, amputation, or even patient death. \g{Iscahemia is characterized by a deficient blood circulation, and it can be recognized by poor vascular reperfusion or dark gangrenous patterns. On the other hand, infection is associated to the presence of tissue inflammation or purulence \cite{goyal_recognition_2020}}. Although continuous monitoring and timely diagnosis is crucial to avoid these serious complications, \g{inappropriate care and lack of screening on diabetic population, as well as} the lack of medical specialists in developing countries poses a barrier towards early discovery and appropriate management of DFU. 
In this setting, \g{computer-supported} diagnosis by means of computer vision techniques and conventional \g{smartphone} cameras represents a promising and economically viable alternative, with some recent research focusing already on developing this kind of technology \cite{yap_computer_2015,yap_new_2018,goyal_recognition_2020}. 
\g{Therefore, previous research has addressed the need of automatic image understanding towards DFU analysis. Ranging from conventional feature engineering-based image classification \cite{veredas_binary_2010} based on colour and texture to more advanced two-stage approaches involving lesion detection or segmentation followed by classification \cite{brungel_detr_2021,amin_integrated_2020}, and more recently deep learning approaches \cite{das_recognition_nodate,rostami_multiclass_2021}. A detailed recent review on machine learning applied to DFU can be found in \cite{tulloch_machine_2020}.}

\begin{figure*}[t]
\centering
\subfloat[]{\includegraphics[width = 0.24\textwidth]{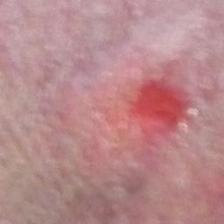}
\label{fig_deg_1}}
\hfil
\subfloat[]{\includegraphics[width = 0.24\textwidth]{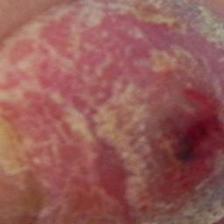}
\label{fig_deg_2}}
\hfil
\subfloat[]{\includegraphics[width = 0.24\textwidth]{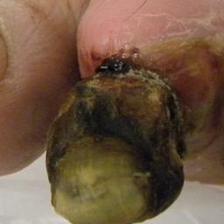}
\label{fig_deg_3}}
\hfil
\subfloat[]{\includegraphics[width = 0.24\textwidth]{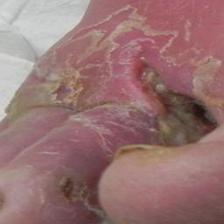}
\label{fig_deg_4}}
\hfil
\caption{Randomly sampled diabetic foot images from the DFUC 2021 dataset. (a) No infection and no ischaemia, (b) presence of infection, (c) presence of ischaemia, (d) presence of both conditions. Images provided by the organization.}
\label{fig_intro}
\vspace{-3mm}
\end{figure*}

Within this context, the \g{Diabetic Foot Ulcer Challenge (DFUC) 2021} was held in conjunction with the MICCAI 2021 conference. 
The task at hand is multi-class classification of foot images into four categories: no pathology, infection, ischaemia, or presence of both pathologies, as illustrated in Fig. \ref{fig_intro}.
This paper describes our approach to solving the challenge, in which we obtained the first position. 
In addition, we make to contributions: 1) we compare the performance of \g{two popular Convolutional Neural Networks (CNNs) \cite{tan_efficientnet_2019,kolesnikov_big_2020} with that of two more recent Vision Transformers \cite{dosovitskiy_image_2021,touvron_training_2021}} in this problem, \g{revealing that CNNs may} still be the preferred choice for this kind of problems, and 2) we experiment with the Sharpness-Aware (SAM, \cite{foret_sharpness-aware_2021}) extension of the popular Stochastic Gradient Descent optimization process, demonstrating that SAM yields superior performance in all the analyzed cases. 
More details on the DFUC 2021 organization, as well as live test leaderboard, can be found online at \url{https://dfu-2021.grand-challenge.org/} . 

\g{In the next section, we first describe the two CNN and the two Transformer architectures we train for solving the task at hand. We then briefly recall the SAM optimization algorithm, and provide a description of the training process. Next, we present the data used for training and validating our analysis, and the performance obtained by each of the considered approaches, as well as the final test performance of our best model. We conclude the paper with some discussion and further research directions.}

\section{Methodology}
We now describe the main methodological aspects of DFU classification in this paper: CNN and Vision Transformer architectures, model optimization process and training data.

\subsection{Convolutional Neural Networks and Vision Transformers}\label{sec_archs}
In this work we experiment with two different kind of neural networks for computer vision, namely conventional Convolutional Neural Networks (CNNs) and more recent Vision Transformers. 
CNNs have become the de-facto approach to \g{most} computer vision tasks in the last decade, since their initial dominance of the 2012 ImageNet challenge \cite{krizhevsky_imagenet_2012}. 
Essential to the learning ability of CNNs from image data are several distinctive features, namely their built-in inductive bias (spatial translation invariance imposed through convolutions with shared learnable filters across the image), or a representation hierarchy arising from their layer-wise architecture design, by which larger patterns are learned as sequential combinations of smaller patterns. 
On the other hand, Transformers have emerged in the last few years as the most powerful architecture for Natural Language Processing applications, dominating virtually all public benchmarks\footnote{Transformers are based on the notion of attention, the detailed description of which falls beyond the scope of this paper \cite{vaswani_attention_2017}.} 
Recently, the Transformer architecture has been extended to other domains and data types, like graphs, speech, or vision \cite{dong_speech-transformer_2018,yun_graph_2019,dosovitskiy_image_2021}. 
After the initial exploration of Transformers for computer vision in \cite{dosovitskiy_image_2021}, where a variety of image recognition tasks were successfully approached with this architecture, the amount of research dedicated to proposing Vision Transformers as an alternative to CNNs has increased exponentially. 
Generally speaking, Vision Transformers consider images as sequences of small patches akin to words or tokens, that can be supplied to standard Transformers. 
However, and in contrast with CNNs, Transformers have no notion of distance within an image, since patches are processed sequentially. 
Spatial relationships need to be modeled by means of positional encodings/embeddings and learned on extremely large datasets, in which case the increased flexibility of Transformers could favor them over CNNs. 
In this paper, we test Transformers' ability to generalize for a relatively small dataset of lesion images like those shown on Fig. \ref{fig_intro}. Specifically, the four architectures we will analyze for our problem are enumerated below.

\begin{itemize}
\item \textbf{Big Image Transfer (BiT)} \cite{kolesnikov_big_2020}. 
We use the ResNeXt50 architecture from this work, which contains 25,55M parameters, uses Group Normalization instead of BatchNorm, and implements convolutional layers with Weight Standardization.
\item \textbf{EfficientNet} \cite{tan_efficientnet_2019}, which is a family of architectures designed by neural architecture search, and carefully scaled so that they achieve state-of-the-art performance with a fraction of parameters. 
In this paper we use the EfficientNet B3 architecture, which contains 12,23M weights.
\item \textbf{Vision Transformers (ViT)} \cite{dosovitskiy_image_2021}, the first architecture to replace convolutions by attention, this is a pure transformer applied on sequences of image patches. 
We use the ViT-base configuration, containing 22,20M weights.
\item \textbf{Data-efficient Image Transformers (DeIT)} \cite{touvron_training_2021}, a refinement of ViT with improved pre-training strategies. 
We use the DeiT-small variant, which contains 22M learnable parameters, and again a patch size of $16\times16$.
\end{itemize}

\subsection{Sharpness-Aware Optimization and Training Details}
In this paper we are also interested in comparing standard Stochastic Gradient Descent (SGD, \cite{bottou_tradeoffs_2008}) to a recently introduced approach for neural network optimization that tries to improve their generalization ability, namely \g{Sharpness-Aware} Optimization (\g{	SAM, \cite{foret_sharpness-aware_2021,korpelevich_extragradient_1976}}) . 
SAM simultaneously minimizes the value of a loss function and the loss sharpness, thereby seeking to find parameters lying in neighborhoods with a uniformly low loss. 
This way, the goal of SAM is to improve model generalization, and it has been shown to deliver State-of-\g{the}-Art performance for different applications, while also enforcing robustness against label noise. 
Further technical details about this optimization algorithm can be found in \cite{foret_sharpness-aware_2021}.

Other than the optimization algorithm choice, all the models trained in this work follow the same process, similarly to \cite{galdran_little_2020}: network weights are optimized so as to minimize the cross-entropy loss with each of the optimizers adopting an initial learning rate of $l=0.01$ and a batch-size of $32$. 
Note that these two values were selected by grid search on an initial train-validation split. 
The learning rate is decayed following a cosine law from its initial value to $l=1e$-$8$ during $3$ epochs, which defines a training cycle. 
This cycle is repeated $25$ times, restarting the learning back at the beginning.
During training images are augmented with standard techniques (random rotations, vertical/horizontal flipping, contrast/saturation/brightness changes). 
The F1-score is monitored on an independent validation set and the best performing model is kept for testing purposes. 
For testing, we generate four different versions of each image by horizontal/vertical flipping, predict on each of them, and average the results.

Note that in all cases, optimization starts from weights pre-trained on the ImageNet classification task\footnote{\g{All architectures and pretrained weights are taken from \cite{wightman_pytorch_2019}}} at a $224\times224$ resolution, which happens to coincide with the size of the image patches provided in the DFUC 2021 challenge, as explained in the next section. 

\g{
\subsection{Training Data}
In DFUC 2021, participants were provided with a dataset composed of 15,683 DFU RGB image patches of a fixed 224$\times 224$ resolution, with 5,955 training and 5,734 testing unlabelled DFU patches for final ranking purposes. An additional set of 3,994 unlabeled DFU patches was released for training, although it was unused in this work.
These patches were extracted from close-ups of diabetic feet at a distance of approximately 30-40 cm perpendicularly to the plane of the ulcer. Natural data augmentation was included (repeated takes of the same lesion), and controlled for with automatic image similarity techniques while splitting the data into training and test subsets. More details on this dataset can be found in \cite{yap_analysis_2021}.
}

\section{Experimental Results and Analysis}
Below we analyze the performance of the four architectures described in section \ref{sec_archs} in the context of the \g{DFUC 2021}. 
The organizers provided participants with a labeled training set and offered an unlabeled validation set \g{as well as} a public leaderboard to which preliminary submissions could be submitted. 
After the validation period was over, the participants had to submit their predictions on a second hidden test set, and the final ranking was based on the latter. 
In all cases, the models described below were trained on a 4-fold split of the labeled training data, with predictions generated by average ensembling.
\g{Models were implemented in PyTorch \cite{paszke_pytorch_2019} and trained on a workstation equipped with a NVIDIA GeForce RTX 3080 with 10 Gb memory size.}

\subsection{Impact of SAM optimization}
In this section we compare the performance of the aforementioned four architectures when trained as specified above, employing both standard SGD and Sharpness-Aware loss minimization with SGD as base optimizer. 
Performance measures include F1-score (the metric that drives the DFU competition), as well as AUC, recall and precision on its macro versions. 
For this experiment, these quantities are measured on the intermediate validation set released by the organizers. Table \ref{tab_results1} shows such results for all cases.

\begin{table*}[!h]  %
	\renewcommand{\arraystretch}{1.4}	
	\centering
\setlength\tabcolsep{8pt}	
\begin{tabular}{l cccc}
 \textbf{Model/Performance}  	  & \textbf{F1-score} & \textbf{AUC}  & \textbf{Recall} & \textbf{Precision} \\
\midrule
\textbf{BiT-ResNeXt50 SGD}        	  & 51.34 & 84.79 &  54.99   & 51.05\\
\textbf{BiT-ResNeXt50 SAM}    	      & 57.71 & 87.68 &  61.88   & 57.74  \\
\hdashline[2pt/5pt]
\textbf{Perf. Diff.}         	  & \textbf{\Plus 6.37} & \textbf{\Plus 2.89} & \textbf{\Plus 6.89}  & \textbf{\Plus 6.69}\\
\midrule 
\textbf{EfficientNet B3 SGD}       & 47.55 & 82.92 &  49.49   &  48.47   \\ 
\textbf{EfficientNet B3 SAM}   	  & 57.65 & 84.81 &  60.74   &  57.13 \\
\hdashline[2pt/5pt]
\textbf{Perf. Diff.}              & \textbf{\Plus 10.05} & \textbf{\Plus 1.89} &  \textbf{\Plus11.25}  & \textbf{\Plus 8.66}\\
\midrule 
\textbf{ViT SGD}       			  & 48.84 & 85.35 & 53.13  & 51.07 \\ 
\textbf{ViT SAM}   			  	  & 51.94 & 87.85 & 56.52  & 54.38 \\
\hdashline[2pt/5pt]
\textbf{Perf. Diff.}              & \textbf{\Plus 3.10} & \textbf{\Plus 2.50} &  \textbf{\Plus3.39} & \textbf{\Plus 3.31} \\
\midrule 
\textbf{DeiT SGD}       		  & 53.87 & 84.92 & 58.01  & 53.01  \\ 
\textbf{DeiT SAM}   			  & 53.98 & 85.97 & 58.12  & 54.03  \\
\hdashline[2pt/5pt]
\textbf{Perf. Diff.}              & \textbf{\Plus 0.11} & \textbf{\Plus 1.05} &  \textbf{\Plus0.11} & \textbf{\Plus 1.02} \\
\bottomrule\\[0.05cm]
\end{tabular}
\caption{Performance analysis of different combinations of architectures with SGD and SAM optimization procedures. \g{Note that all metrics are macro-averaged across categories.}}
\label{tab_results1}
\end{table*}%

\subsection{CNNs vs Transformers}
We also study the performance of each of our four architectures in the final hidden test set. 
For this, Table \ref{tab_results2} shows the metrics obtained by each model, whereas Table \ref{tab_results3} describes the result of our final submission for the challenge classification, compared to the results obtained by the top-ranked contestants.

\begin{table*}[!h]  %
	\renewcommand{\arraystretch}{1.6}	
	\centering
\setlength\tabcolsep{8pt}	
\begin{tabular}{l cccc}
 \textbf{Model/Performance}  	      & \textbf{F1-score} & \textbf{AUC}  & \textbf{Recall} & \textbf{Precision} \\
\midrule
\textbf{BiT-ResNeXt50}     			      & \textbf{61.53}    & \textbf{88.49}&  \textbf{65.59} & \textbf{60.53}  \\
\textbf{EfficientNet B3}	  		  & 59.71             & 87.01 		  &  61.79 			& 59.34  \\
\textbf{ViT} 		   	      	      & 58.48             & 87.64 		  &  62.27   		& 58.91  \\
\textbf{DeiT}    	      		      & 57.29             & 87.98 		  &  61.35   		& 57.01  \\
\bottomrule\\[0.05cm]
\end{tabular}
\caption{Performance analysis in the final test set for the four different architectures considered in our comparison. \g{Note that all metrics are macro-averaged across categories.}}
\label{tab_results2}
\end{table*}%

\begin{table*}[t]  %
	\renewcommand{\arraystretch}{1.6}	
	\centering
\setlength\tabcolsep{8pt}	

\begin{tabular}{l cccc}
 \textbf{Participant/Performance} & \textbf{F1-score} & \textbf{AUC}  & \textbf{Recall} & \textbf{Precision} \\
\midrule
\textbf{agaldran (this work)}     & \textbf{62.16}    & \textbf{88.55}&  \textbf{65.22} & 61.40\\
\textbf{Louise.Bloch}	  		  & 60.77             & 86.16 		  &  62.46 			& \textbf{62.07}  \\
\textbf{S. Ahmed}    	      	  & 59.59             & 86.44 		  &  59.79   		& 59.84  \\
\textbf{Abdul}    	      		  & 56.91             & 84.88 		  &  61.04   		& 58.14  \\
\textbf{orhunguley}       		  & 56.10             & 87.02 		  &  57.59   		& 59.17  \\
\bottomrule\\[0.05cm]
\end{tabular}
\caption{Performance analysis in the final test set for the five top competitors \g{sorted by ranking in terms of F1-score}. Note that our submission in this table corresponds to an ensemble of the predictions from the BiT-ResNeXt50 and EfficientNet B3 models in Table \ref{tab_results2}. \g{Note that all metrics are macro-averaged across categories.}}
\label{tab_results3}
\end{table*}%

\subsection{Discussion of the Results}
The first conclusion we can draw from the above results is that the SAM optimization generally benefits the end performance of both CNNs and Transformers, with CNNs always appearing to be superior to Transformers in all cases.
It should be noted that SAM requires two forward passes instead of one before weight update, which turns it into a relatively expensive alternative to SGD. 
However, results in Table \ref{tab_results1} appear to indicate that the extra computational load is widely compensated, particularly when working with small datasets for which the cost of one training run is not \g{substantial}.

Next, we can see in Table \ref{tab_results2} that, as expected, the relative ranking of our four models in the validation set is preserved when evaluating them in the final test set.
However, there is a noticeable increase in the performance of Vision Transformers, which attain results close to those of the EfficientNet B3 architecture. 
It is also worth stressing that, just as in the validation experiments on Table \ref{tab_results1}, the BiT-ResneXt50 model still achieves the highest performance, surpassing by a good margin the second best model, which is EfficientNet B3.

Finally, we see from the final ranking of the DFUC 2021 challenge\footnote{Ongoing (live) leaderboard can be accessed at \url{https://dfu-2021.grand-challenge.org/evaluation/live-testing-leaderboard/leaderboard/}.} in Table \ref{tab_results3} that our approach is validated when compared to other contestants' results. 
Our final highest score was achieved by a linear combination of the predictions extracted from BiT-ResneXt50 and EfficientNet B3, but it is worth noting that BiT-ResneXt50 alone would still be the ranked the best method, and EfficientNet B3 would have been ranked second.

\section{Conclusion}
This paper describes the winning solution to the DFUC 2021 grand-challenge on Diabetic Foot Ulcer Classification, held in conjunction with MICCAI 2021, which is based on an ensemble of CNNs (BiT-Resnext50 and EfficientNet B3) trained on different data folds.
In addition, for weight optimization we employ Sharpness-Aware Minimization, which provides a noticeable improvement in performance for all considered models. 
\g{Note that, in the literature, SAM has been benchmarked mostly with CNNs, which could introduce a bias of this optimization algorithm towards that option in this paper. 
However, it has been recently shown that SAM has equally good impact in performance for Vision transformers \cite{chen_when_2021}.}
Finally, we also include a comparison of the performance of CNNs and Vision Transformers for this task, which reveals that despite the recent popularity of Transformers for Computer Vision tasks, CNNs may still be the preferred choice in few-data scenarios.

\section*{Acknowledgments}
Adrian Galdran was funded by a Marie Skłodowska-Curie Global Fellowship (No 892297). 
Gustavo Carneiro was partially supported by Australian Research Council grants (DP180103232 and FT190100525)

\bibliographystyle{splncs04}
\bibliography{dfu_challenge.bib}

\end{document}